\documentclass[a4paper,fleqn,usenatbib]{mnras}
\usepackage[T1]{fontenc}
\usepackage{ae,aecompl}
\usepackage{subfig}
\usepackage{graphicx}
\usepackage{caption}
\usepackage{float}
\usepackage{amsmath}
\usepackage{amssymb}
\usepackage{footnote}

\title[$\langle \Pi \rangle$ polarisation of radio sources in \textit{Planck}]
{Statistics of the fractional polarisation of compact radio sources in \textit{Planck} maps}

\author[L. Bonavera]{
Laura Bonavera,$^{1}$\thanks{E-mail: laurabonavera@gmail.com}
Joaquin Gonz\'{a}lez-Nuevo,$^{1}$
Francisco Arg\"ueso,$^{2}$
Luigi  Toffolatti$^{1,3}$
\\
$^{1}$Departamento de F\'{i}sica, Universidad de Oviedo, C. Federico Garc\'{i}a Lorca 18, 33007 Oviedo, Spain\\
$^{2}$Departamento de Matem\'aticas, Universidad de Oviedo, C. Federico Garc\'{i}a Lorca 18, 33007 Oviedo, Spain\\
$^{3}$INAF/IASF Bologna, Via Gobetti 101, 40129 Bologna, Italy\\
}

\date{Accepted XXX. Received YYY; in original form ZZZ}

\pubyear{2016}

\begin{document}
\label{firstpage}
\pagerange{\pageref{firstpage}--\pageref{lastpage}}
\maketitle

\begin{abstract}
In this work we apply the stacking technique to estimate the average fractional polarisation from 30 to 353 GHz of a primary sample of 1560 compact sources  -- essentially all radio sources -- detected in the 30 GHz \textit{Planck} all-sky map and listed in the second version of the \textit{Planck} Catalogue of Compact Sources (PCCS2). We divide our primary sample in two subsamples according to whether the sources lay (679 sources) or not (881 sources) inside the sky region defined by the \textit{Planck} Galactic mask ($f_{\rm sky}$ $\sim 60$ {\it per cent}) and the area around the Magellanic Clouds.
We find that the average fractional polarisation of compact sources is approximately constant (with frequency) in both samples (with a weighted mean over all the channels of 3.08 {\it per cent} outside and 3.54 {\it per cent} inside the \textit{Planck} mask).
In the sky region outside the adopted mask, we also estimate the $\mu$ and $\sigma$ parameters for the log-normal distribution of the fractional polarisation, finding a weighted mean value over all the {\it Planck} frequency range of 1.0 for $\sigma$ and 0.7 for $\mu$ (that would imply a weighted mean value for the median fractional polarisation of 1.9 {\it per cent}).
\end{abstract}




\section{Introduction}

The polarisation properties of extragalactic radio sources (ERS) at frequencies above 10--20 GHz are still poorly
constrained by observations. The NRAO Very Large Array (VLA) Sky Survey \citep[NVSS at 1.4 GHz][]{CON98} still constitutes the largest sample of ERS
surveyed both in total flux density, $S$, and in total linear polarisation, $P$. More recently, the Australia Telescope Compact Array (ATCA), the VLA and other facilities have been extensively used to observe and characterize the polarization properties of ERS sources up to $\sim 40$ GHz. However, extrapolations from these low
frequencies up to $\sim 100$ GHz are affected by large uncertainties since a complex combination of effects must be considered. This
includes intra-beam effects and bandwidth depolarisation, in addition to possible intrinsic frequency-dependent
changes \citep[see][for comprehensive discussions on this subject]{TUC04,TUC12}.

Moreover, extending the knowledge of the polarisation properties of ERS is interesting on its own, since it
provides information about the physics of the underlying emission process. Thus, to fill this gap of knowledge, in
the past decade many intermediate to large samples of ERS have been observed in polarisation at frequencies above
$\simeq 10$ GHz \citep{SAD06,MAS08,MAS11,MAS13,LOP09,JAC10,MUR10,SAJ11}. All these surveys have provided values of the median fractional polarisation of ERS in the range $\sim 2$ per cent to $\sim 3$ per cent of the total flux density of the source.
Other recent studies of the polarized emission in ERS have tried to analyse the dependence of the fractional
polarisation with luminosity, redshift, and the source environment. The current, still preliminary, results show
no correlation between the fractional polarisation and redshift, whereas a weak correlation is found between
decreasing luminosity and increasing degree of polarisation \citep[see, e.g.,][]{BAN11}.

As reminded above, the total polarisation, $P$, commonly observed in ERS at cm or mm wavelengths is typically a
few percent with only very few ERS showing a total fractional polarisation, $\Pi= P/S$, as high as $\sim 10$ per cent of
the total flux density \citep{SAJ11,TUC12}. However, even this low $\Pi$ of ERS may constitute a problem for the detection of the primordial polarisation in the Cosmic Microwave Background (CMB) maps, since ERS are the dominant polarized foreground at small angular scales and the CMB polarized signal constitutes only a few percent \citep[e.g.,][]{PLA1511,PLA15par} of the CMB temperature anisotropy. This very low intrinsic CMB polarized signal means that current experiments are marginally able to detect tensor metric perturbations by the direct detection of the primordial CMB B-mode, but only in the case of very high values of the tensor-to-scalar ratio of primordial perturbations, $r \sim 0.05-0.1$. For this reason, with the most recent available data, only upper limits on the value of $r$ have been obtained \citep{BIC15,KEC16}.

Therefore, future CMB all-sky surveys in polarisation \citep[e.g., the proposed European Space Agency, ESA, Cosmic ORigin Explorer, CORE, mission, see][a mission specifically designed to detect CMB polarisation by virtue of a much higher sensitivity than before]{CORE} -- with the capability to reach tensor-to-scalar ratios as low as $r\sim 0.01$ -- will surely need a more careful determination of the polarized signal from the (dominant) Galactic diffuse emission but also of the signal coming from ERS  \citep[see][for comprehensive reviews on this latter subject]{DEZ15,DEZ16}. To achieve this goal, a much better characterization of the contaminating signal due to polarized ERS in CMB anisotropy maps is needed. And this characterization will be only possible by a proper statistical knowledge of the polarisation properties of the populations of faint compact sources at mm wavelengths.

This is not an easy task since the current available all--sky catalogues of ERS at mm/sub-mm wavelengths are still limited to the shallow surveys provided by the Wilkinson Microwave Anisotropy Probe (WMAP) \citep{BEN03} and ESA \textit{Planck} \citep{PLA15gen} missions. The second, updated, version of the \textit{Planck} Catalogue of Compact Sources (PCCS2) \citep{PLA15pccs2} currently constitutes the deepest complete catalogue of compact radio sources at high frequency ($> 30$ GHz) but, due to the low resolution of the \textit{Planck} beams, it is still limited to very bright sources ($S> 100$ mJy, at the 90 per cent completeness level), even in the cleanest \textit{Planck} channels. Correspondingly, the number of detected compact sources in polarisation is very low (only few tens are detected) and their polarisation properties are poorly characterized.
Moreover, it is obvious that only compact -- either Galactic or extragalactic -- sources with a high $\Pi$ can be detected and, thus, the statistical characterization of the underlying population will be biased
towards these highly polarized objects. The use of samples of detected compact sources also hampers the possibility
of studying the possible variation of $\Pi$ with the total flux density \citep{MAS13}. In particular, the source populations mainly contributing to the source counts down to very faint fluxes could present different $\Pi$ values as compared to the ones dominating the bright number counts \citep[e.g.,][]{TUC04}.

To reduce, albeit partially, this lack of data on faint, undetected, compact sources and for better estimating the
average fractional polarisation, $\langle \Pi \rangle$, of the underlying source population, it is useful to exploit the full
information content of CMB sky maps in polarisation by applying stacking techniques, i.e., co-adding the signal
from many weak or undetected objects to obtain a statistical detection. Stacking has been already successfully
applied to \textit{Planck} data to search for the ISW signal directly at the positions of positive and/or negative
peaks in the gravitational potential, since the expected (and observed) signal is very weak \citep{PLAisw}. 
Moreover, stacking has also been successfully applied to investigate the faint polarized signal of radio sources, detected in total flux density, $S$, at 1.4 GHz by NVSS. By stacking the polarized signal for NVSS sources down to the detection limit in total flux density the authors \citep{STI14} were able to find a gradual increase -- with decreasing flux density -- in the median fractional polarisation fully consistent with a trend noticed before for bright NVSS sources \citep{MES02,TUC04}.

The outline of the paper is as follows: in Section 2 we discuss the methods adopted for selecting the sky patches, for defining the sub-samples we are going to analyse and for determining their mean fractional polarisation; in Section 3 we present our results on the fractional
polarisations of compact sources, inside and outside the \textit{Planck} mask, and estimated by stacking; finally,
in Section 4, we summarize our main conclusions.

\section{Methods}

\begin{figure}
 \centering
  \subfloat{\includegraphics[width=9.0cm]{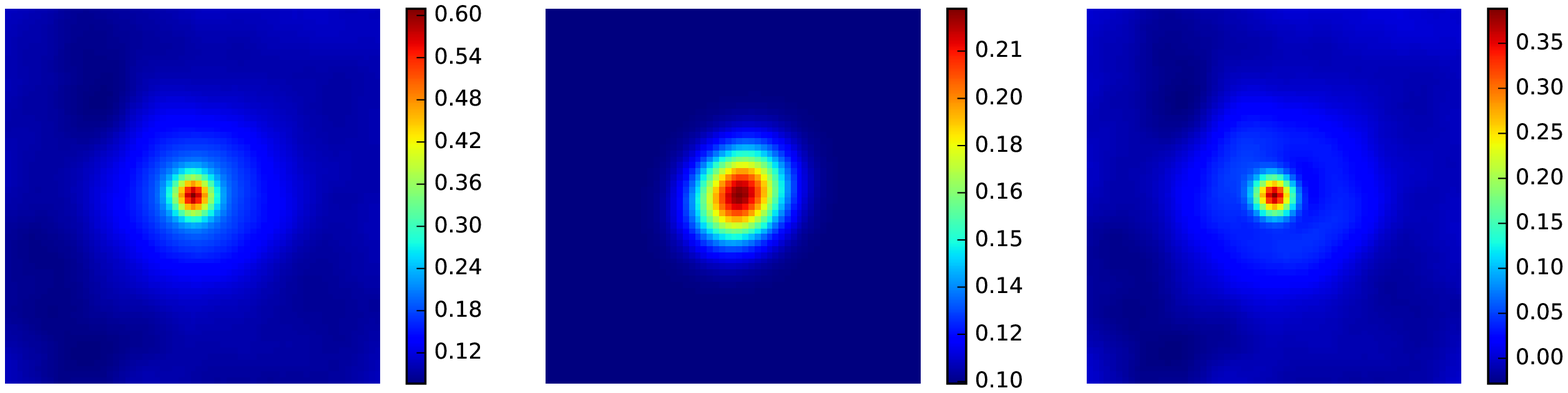}}\\
  \subfloat{\includegraphics[width=9.0cm]{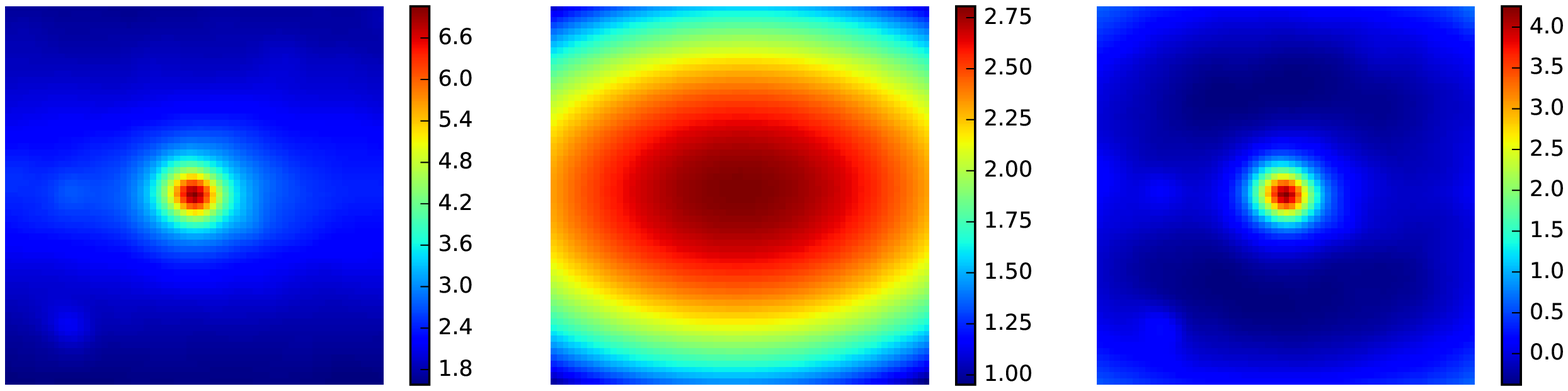}}\\
   \caption{From left to right: mean maps of the original S data, the model fit to the background (see text) and the ``residual" for the 143 GHz channel, outside (top) and inside (bottom) the adopted {\it Planck} mask. The color scale is Jy/pix, the pixel size is 1.72' and the angular size of the image is $3.14$ $deg^2$. The values in this figure are not yet corrected for the noise bias.}
 \label{fig:gaufit}
\end{figure}  

\subsection{Data}
\label{sec:data} 

Our purpose is to study the main behaviour of compact radio sources in polarisation at high frequencies. For this reason we use the PCCS2 \citep{PLA15pccs2} catalogue at 30 GHz\footnote{Based on observations obtained with {\it Planck} (http://www.esa.int/Planck), an ESA science mission with instruments and contributions directly funded by ESA Member States, NASA, and Canada. Available at http://pla.esac.esa.int/pla/\#home (\textit{Planck} Legacy Archive).} as our main sample. It consists of 1560 sources above a flux density of 427 mJy at the 90 per cent completeness level (and with a minimum flux of 376 mJy).

We follow these sources in the \textit{Planck} maps at 30, 44, 70, 100, 143, 217 and 353 GHz\footnotemark[\value{footnote}]  \citep{PLA15gen}, that corresponds to the channels with both total intensity and polarisation measurements. Since we are interested in estimating $\langle \Pi \rangle$ and $\langle \Pi^2 \rangle$, we generate the P map as the square root of the quadratic sum of the Stokes Q and U parameters maps ($P=\sqrt{Q^2+U^2}$).

Due to the different behaviour of sources inside and outside the Galactic plane in this wide range of frequencies, we decide to split our sample into two subsamples according to  the regions delimited by the \textit{Planck} Galactic mask GAL060 that leaves 60 per cent of the sky unmasked\footnotemark[\value{footnote}]. We also include in the Galactic region a circular sky region of $5$ $deg$ radius around the position of the Large Magellanic Cloud and of $3$ $deg$ radius around the Small Magellanic Cloud.

\begin{table*}
 \centering
\begin{tabular}{ l | c | c | c | c  | c  | c | c | c | c | c | c | r } 
\hline
&  \multicolumn{12}{c}{Extragalactic region} \\
 &  \multicolumn{2}{c}{Uncorrected} & \multicolumn{2}{c}{Corrected} &  \multicolumn{2}{c}{Uncorrected} & \multicolumn{2}{c}{Corrected} & \multicolumn{4}{c}{Log-normal parameters} \\
freq &  $\langle \Pi \rangle$ & error &  $\langle \Pi \rangle$ & error & $ \sqrt{\langle \Pi ^2 \rangle}$ & error & $ \sqrt{\langle \Pi ^2 \rangle}$ & error & $\mu$ & error & $\sigma$ & error \\
$[GHz]$  &per cent & per cent & per cent & per cent  & per cent & per cent  & per cent & per cent & & &  &\\
\hline
30   & 0.65 & 0.11 & 3.05 & 0.21 & 3.41 & 0.22 & 4.03 & 0.65 & 0.8  & 0.2 & 0.7 & 0.2\\ 
44   & 0.64 & 0.23 & 3.27 & 0.55 & 4.66 & 0.80 & 4.64 & 1.58 & 0.8  & 0.5 & 0.8 & 0.4\\ 
70   & 0.44 & 0.26 & 2.51 & 0.48 & 4.47 & 0.58 & 4.48 & 0.49 & 0.3  & 0.4 & 1.1 & 0.2\\ 
100 & 0.91 & 0.19 & 3.26 & 0.28 & 4.97 & 0.29 & 5.75 & 0.47 & 0.6  & 0.2 & 1.1 & 0.1\\ 
143 & 1.05 & 0.20 & 3.06 & 0.28 & 4.33 & 0.18 & 4.91 & 0.51 & 0.6  & 0.2 & 1.0 & 0.1\\ 
217 & 1.09 & 0.25 & 3.07 & 0.29 & 3.79 & 0.34 & 4.54 & 0.41 & 0.7  & 0.2 & 0.9 & 0.1\\ 
353 & 0.99 & 0.60 & 3.52 & 1.20 & 3.78 & 1.21 & 4.64 & 0.94 & 1.0  & 0.7 & 0.7 & 0.5\\ 
\hline
\end{tabular} 
 \caption{From left to right: frequency, mean fractional polarisation with r.m.s. errors uncorrected and corrected for the noise bias, square root of the mean quadratic fractional polarisation with 1$\sigma$ errors uncorrected and corrected for the noise bias, $\mu$ and $\sigma$ parameters of the log-normal function characterizing the mean fractional polarisation distributions (see text) and their 1$\sigma$ errors. The results are for the case outside the \textit{Planck} Galactic mask with $f_{\rm sky}$ = 60 per cent and the background fit as described in the text.}
   \label{tab:polfrac_ext}
 \end{table*}

\begin{table}
 \centering
\begin{tabular}{ l | c | c | c | r } 
\hline
&  \multicolumn{4}{c}{Galactic region} \\
 &  \multicolumn{2}{c}{Uncorrected} & \multicolumn{2}{c}{Corrected} \\
freq &  $\langle \Pi \rangle$ & error &  $\langle \Pi \rangle$ & error \\
$[GHz]$  &per cent & per cent & per cent & per cent  \\
\hline
30   & 3.23 & 0.60 & 7.30 & 0.72 \\ 
44   & 0.74 & 0.09 & 3.67 & 0.13 \\ 
70   & 1.09 & 0.06 & 3.74 & 0.13 \\ 
100 & 1.35 & 0.22 & 3.88 & 0.24 \\ 
143 & 1.23 & 0.10 & 3.11 & 0.12 \\ 
217 & 1.16 & 0.19 & 3.05 & 0.14 \\ 
353 & 1.42 & 0.27 & 4.30 & 0.19 \\ 
\hline
\end{tabular} 
 \caption{From left to right: frequency, mean fractional polarisation with 1$\sigma$ errors uncorrected and corrected for the noise bias. The results are for the case inside the \textit{Planck} Galactic mask with $f_{\rm sky}$ = 60 per cent and the background fit as described in the text.}
   \label{tab:polfrac_gal}
 \end{table}

\subsection{Stacking}
\label{sec:stacking} 

Stacking is a statistical method that consists in adding up many regions of the sky around previously selected positions \citep[see][and references therein]{DOL06,MAR09,BET12}. In this way we can reduce the noise/background, since it is expected to fluctuate around the mean with positive and negative values, and enhance the signal we want to study.
Stacking is useful when the individual sources in the selected sample are too faint to be detected at the frequency of interest. Therefore we can measure their mean flux density despite the fact that they are not detectable directly by the given instrument. 

In our case we want to perform statistical estimates of polarization with \textit{Planck}. It should be noticed that our target sources are all detected in total intensity at 30 GHz, but they are not necessarily detected in the higher frequency channels: their spectral behaviour is, in fact, typically flat with down-turning spectra above 70 GHz \citep{PLAstat,PLA13pccs}. In polarisation the situation gets worse since only few tens of sources are detected even at 30 GHz \citep{PLA15pccs2}. For this reason our scientific case is an excellent application of the stacking methodology.

In particular, with stacking in total intensity and polarisation we compute the $\langle \Pi \rangle$ of our source population at frequencies up to 353 GHz (all the \textit{Planck} channels with polarisation measurements).
To perform stacking we select a small patch of 63 x 63 pixels (corresponding to $\sim40$ times the solid angle of the beam at 30 GHz) around each source position. The pixel size is 3.44' for LFI and 1.72'  for HFI channels. We then add up all the patches to obtain the total flux density. To reduce the instrumental noise (a second order effect) we convolve the resulting patch  with a Gaussian filter whose $\sigma_{\rm filter}$ is given by $\sigma_{\rm beam}/2$.

Unlike the case for \cite{STI14}, where they use high resolution radio data at 1.4 GHz, in the microwave band and at the much lower {\it Planck} angular resolution we need to take into account some additional signals that are plausible contaminants to our stacked measurements: the CMB itself and the diffuse emission of our own Galaxy. Although their contribution is very small in polarisation, it is not negligible when stacking over hundreds of targets. In the final stacked image these small contributions give rise to a strong background signal that has to be removed. Therefore, we estimate and subtract it from the final stacked patch, where the contamination signal can be more easily calculated both in total intensity and in polarisation.
In polarisation we subtract the mean of the background computed in the external region of the final patch (3$\sigma_{\rm beam}$ away from the patch center) from the total polarisation flux.

In the subsample outside the Galactic mask region, the sources lying on a CMB maximum are more likely to be detected, due to the bias in the source detection in total intensity. In the HFI channels (where the S/N is higher) this fact results in an extended feature around the center of the final patch. In this latter case, we estimate the background by fitting it with a constant plus Gaussian 2D curve and then subtract it from the data.  This effect is clearly not present either in polarisation or in our injected sources test (see Section \ref{sec:s_injection}). 

For the subsample inside the Galactic region, due to the Galactic emission gradient, we need to perform a parabolic 2D fit to more carefully estimate the background.
We also compute the flux densities in total intensity with the same background subtraction procedure used for polarisation and compare both sets of results in Section \ref{sec:results}.
From these residual maps we then compute the total flux densities in total intensity and polarisation.

An example of background estimation is illustrated in Fig. \ref{fig:gaufit}. It refers to the 143 GHz case. On the left, the total intensity map resulting from the stacking in the region outside (top) and inside (bottom)  the Galactic mask is shown. The central panel is the model obtained by fitting the background in total intensity. In the top figure, the fit is performed using a flat component and a 2D Gaussian curve to take into account the noisy background and the contribution from the positive CMB fluctuations.
In the bottom panel we use a 2D, two degree polynomial function to model the background in the Galactic region. The right panels show the residual maps which we use to estimate the flux density.
It is worth mentioning that in the test with injected sources (described in Section \ref{sec:s_injection}) there was no need to perform such background fitting, because the boosting effect was not present, strengthening the idea that it is due to a detection bias. It should also be stressed that the values in this figure are not corrected for the noise bias (see Section \ref{sec:noisebias}).

Finally we compute $\langle \Pi \rangle = \langle P_0 \rangle / \langle S \rangle$, where $P_0$ is the source total polarisation amplitude and $\langle P_0 \rangle$ its average over our sample. Its error is given by 
$ \sqrt{ \left( \langle P_0 \rangle / \langle S \rangle \right)^{2}  \cdot \left(  \sigma_{P_0}^2 /  \langle P_0 \rangle ^2 + \sigma_S ^2/  \langle S \rangle ^2 \right) }$,
 where $\sigma_{P_0}$ and $\sigma_S$ are the standard deviations for total intensity and polarisation computed in the external region of the stacked patches. 

We also compute the quantity $\sqrt{\langle \Pi^2 \rangle} = \sqrt{\langle P_0^2 \rangle / \langle S^2 \rangle}$ by applying the same methodology with the only difference that due to the higher S/N we don't need to perform any background fit to compute $\sqrt{\langle \Pi^2 \rangle}$. Its error is given by $\sqrt{\dfrac{1}{4 \langle P_0^2 \rangle \langle S^2 \rangle}  \left(   \sigma^2_{P_0^2} + \langle P_0^2 \rangle^2/ \langle S^2 \rangle^2 \sigma^2_{S^2} \right)  }   $.

Please note that most of our sources are not directly detectable and therefore we cannot estimate directly $\langle \Pi \rangle=\langle P_0 /S \rangle$ and $\langle \Pi^2 \rangle=\langle P_0^2 /S^2 \rangle$. For this reason, in our stacking procedure we decided to calculate $\langle \Pi \rangle=\langle P_0 \rangle / \langle S \rangle$ and $\langle \Pi^2 \rangle=\langle P_0^2 \rangle/\langle S^2 \rangle$ that are good approximations for $\langle \Pi \rangle=\langle P_0 /S \rangle$ and $\langle \Pi^2 \rangle=\langle P_0^2 /S^2 \rangle$, taking into account that $\Pi$ and $S$ can be considered independent variables (see e.g. \cite{MAS08, MAS13} and \cite{GAL17} as discussed in Section \ref{sec:comparison}). Besides, the residual errors introduced by these assumptions are much lower than the noise bias discussed in Section \ref{sec:noisebias}. Moreover, the bias subtraction methodology described in Section \ref{sec:s_injection} also corrects any residual deviation  from the theoretical value.

\begin{figure*}
 \centering
 \includegraphics[width=16.0cm]{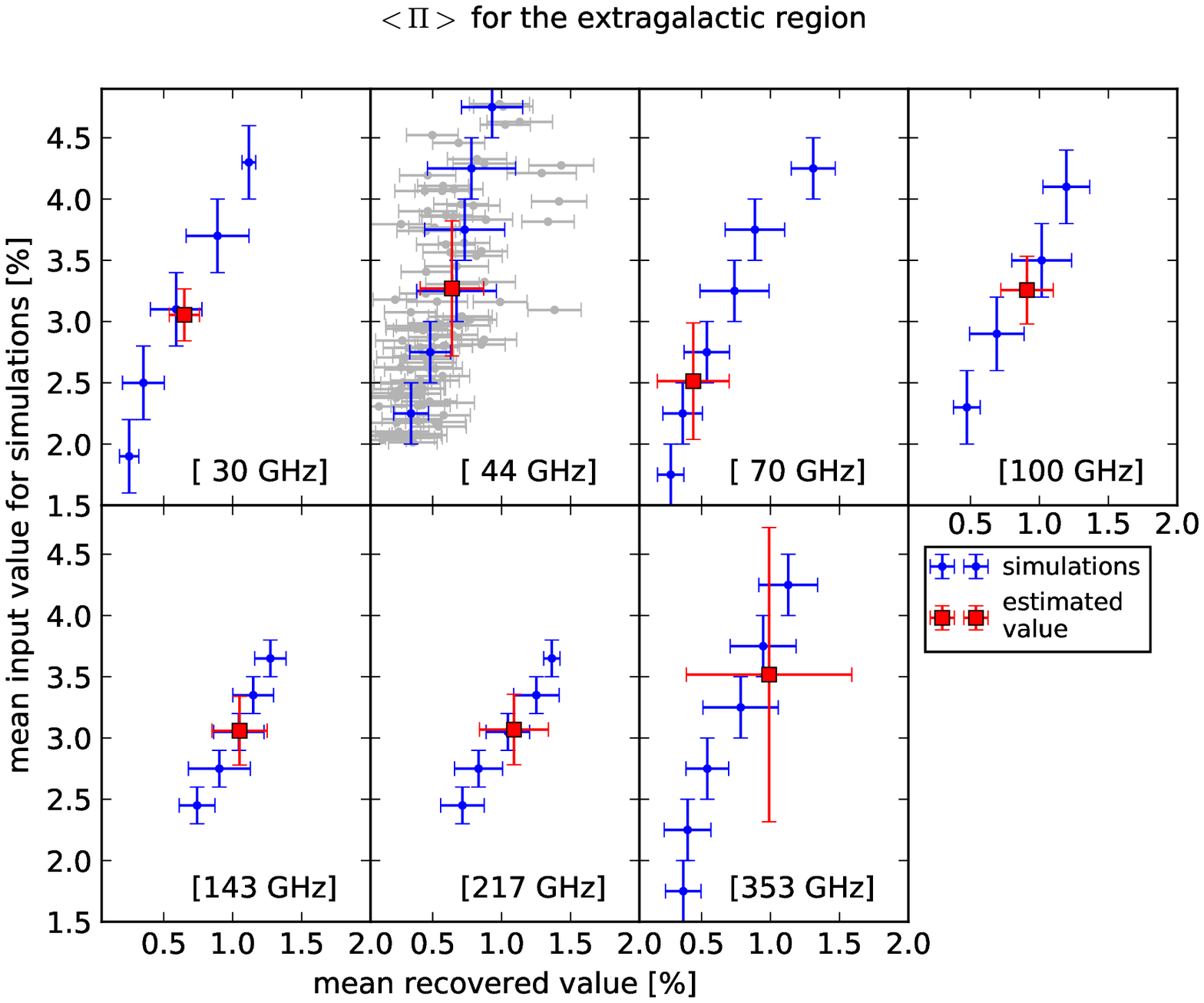}
 \caption{Results obtained outside the adopted mask ({\it extragalactic region} of the sky) from 30 to 353 GHz. As an example, in the 44 GHz panel the grey points are obtained with each individual simulation: on the y-axis we plot the mean input $\left\langle \Pi \right\rangle $ value for simulations and the x-axis is the value recovered with stacking for different values of $\mu$ and $\sigma$, as described in the text. For all the panels, the linear interpolation of these points gives us the correction for the noise bias that has to be applied to the observed values (red squares). The blue points are obtained by averaging over the simulations points with a binning step of about 0.5 in the y-axis. }
 \label{fig:ex_res}
\end{figure*}

\begin{figure*}
 \centering
 \includegraphics[width=16.0cm]{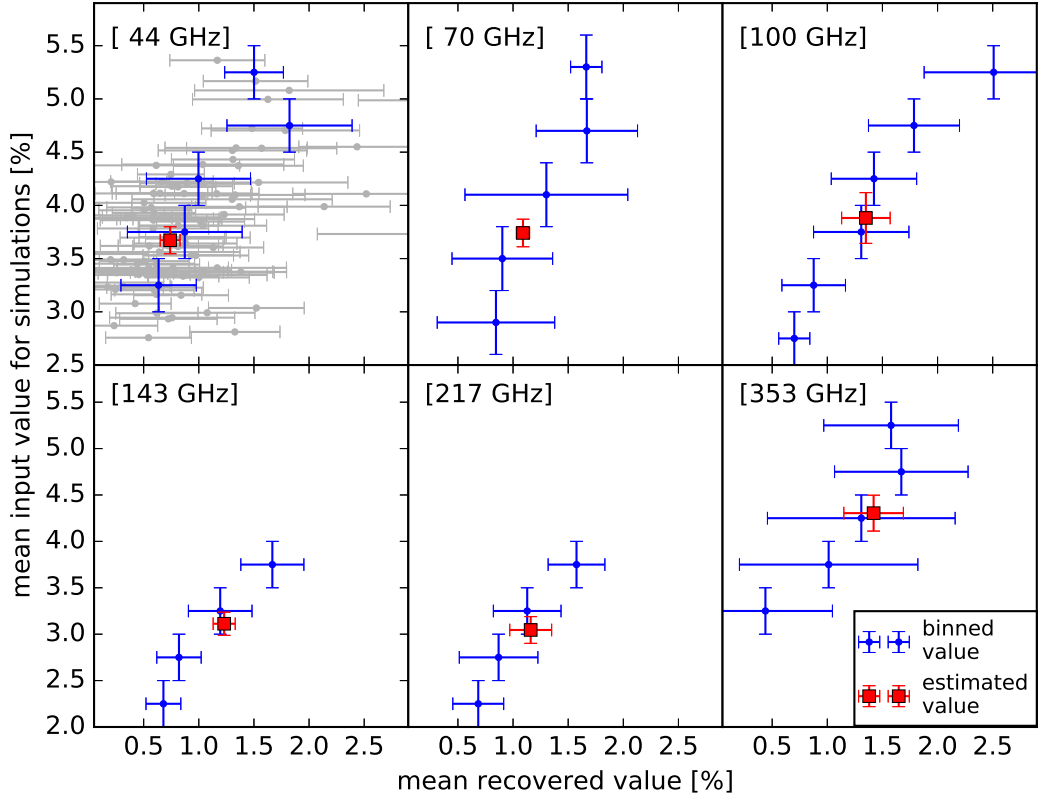}
 \caption{Results obtained inside the adopted mask ({\it Galactic region} of the sky) from 44 to 353 GHz. As in Fig.\ref{fig:ex_res}, the grey points in the 44 GHz panel are obtained with each individual simulation. The red squares are the values obtained with stacking on our sample's positions and corrected for the noise bias using the linear interpolation and the blue points are the binned version of the simulations points.}
 \label{fig:gal_res}
\end{figure*}

\begin{figure}
 \centering
 \includegraphics[width=9.0cm]{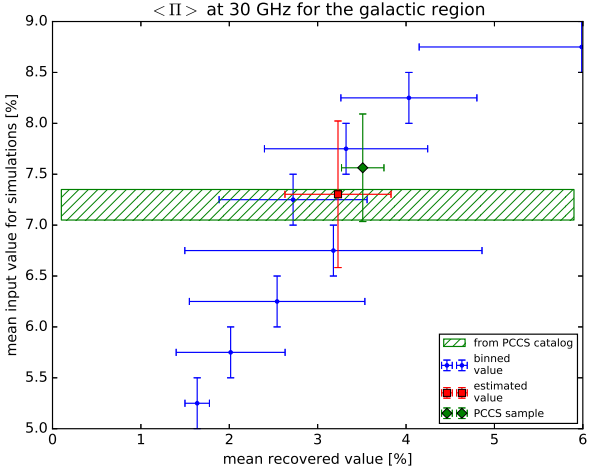}
 \caption{ {\it Galactic region} results at 30 GHz. The blue points are from the binned version of the simulations results: input $\left\langle \Pi \right\rangle $ values on the y-axis and recovered $\left\langle \Pi \right\rangle $ values with stacking on the x-axis. The red square is the value obtained with stacking on our Galactic sample positions (x-axis) and the corrected value (y-axis), according to the linear interpolation. The green diamond is the result obtained with the subsample $P > 2P_{\rm err}$ in the PCCS2 catalogue to be compared with the green hatched stripe (actual value computed with the PCCS2 flux densities for the same subsample).}
 \label{fig:gal_res30}
\end{figure}

\begin{figure*}
 \centering
 \includegraphics[width=16.0cm]{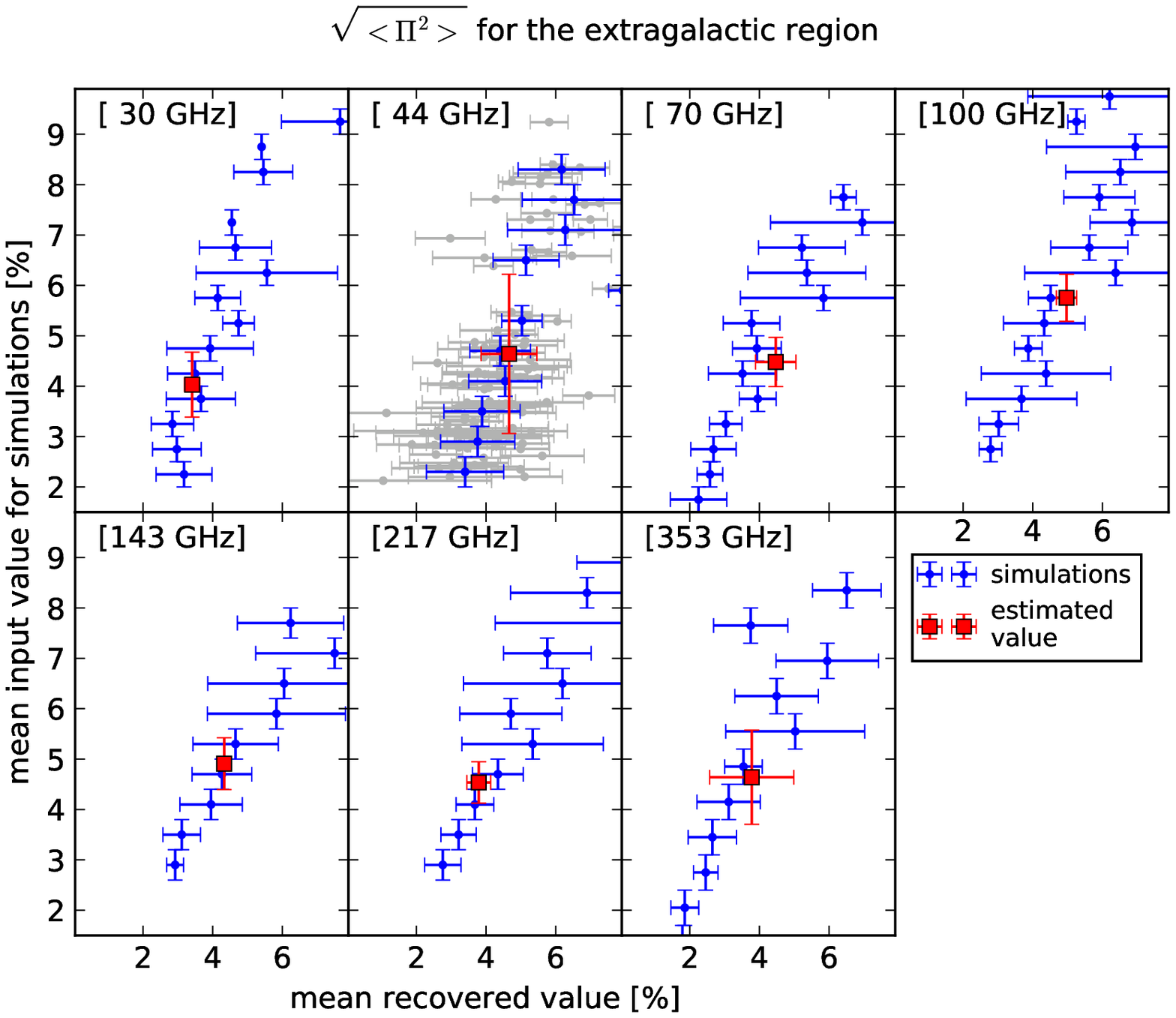}
 \caption{Results obtained for $\sqrt{\left\langle \Pi^2 \right\rangle} $  outside the adopted mask ({\it extragalactic region} of the sky) from 30 to 353 GHz. As an example, in the 44 GHz panel the grey points are obtained with each individual simulation: the y-axis is the mean input $\sqrt{\left\langle \Pi^2 \right\rangle} $  for simulations and the x-axis is the value recovered with stacking for different values of $\mu$ and $\sigma$, as described in the text. For all the panels, the linear interpolation of these points gives us the correction for the noise bias that has to be applied to the observed values (red squares). The blue points are obtained by averaging over the simulations points with a binning step of about 0.5 in the y-axis.}
 \label{fig:ex_res_pi2}
\end{figure*}

\subsection{Noise bias}
\label{sec:noisebias} 

As stated in Section \ref{sec:data}, $P$ is a quadratic sum of $Q$ and $U$. This construction introduces a bias due to the $Q$ and $U$ noise cross-terms. This bias is usually referred as noise bias and it is not negligible for low S/N observations. For this reason it has to be taken into account in order to correct the measurements. Different methods have been developed for this correction \citep[see, e.g.,][]{SIM85,VID16}, but none of them can be directly applied to our case. In fact, in our approach it is important to keep in mind that - by stacking - we are neither observing nor detecting individual sources. This is due to the fact that the polarisation signal-to-noise ratio is not high enough to guarantee their direct detection. Moreover the noise bias strongly depends on the S/N that is unknown in our case for each individual compact source. Therefore, in order to get a theoretical estimation of the importance of the noise bias in our procedure, we proceed by considering the $n$ sky patches constructed around the sources detected in the intensity maps and by adding them up in polarisation to increase the signal-to noise ratio (stacking method).

As previously discussed, these sources have not been individually detected in the polarisation maps. However, we assume that each of these sources is polarized with total polarisation amplitude $P_0=\sqrt{Q_0^2+U_0^2}$, being $Q_0$ and $U_0$ the polarisation amplitudes in $Q$ and $U$, respectively. Far from the centre of the patch- a $3\sigma_{beam}$ distance is enough- we assume that the distribution of the values of $Q$ and $U$ is Gaussian with zero mean and r.m.s deviation $\sigma=\sigma_Q=\sigma_U$, with $\sigma$  coming from the combination of instrumental noise, CMB and the different foregrounds.

Therefore, in one observed sky patch with no compact sources the polarisation $P=\sqrt{Q^2+U^2}$ follows a Rayleigh distribution \citep{PAP84} in the pixels outside the source:
\begin{equation}
 f(P)=P/\sigma^2\, \exp{(-P^2/2\sigma^2)} 
\end{equation} 
\noindent On the contrary, if in the central pixel there is a polarized source with total polarisation amplitude $P_0$, we are left with a Rice distribution in P \citep{RIC54}
 \begin{equation} 
 f(P)=P/\sigma^2\, \exp{(-(P^2+P_0^2)/2\sigma^2)} \, I_0(P_0P/\sigma^2)  
\end{equation}
\noindent with $I_0$ the modified Bessel function of order zero. The expectation of $P$ is for the Rayleigh distribution (absence of sources)
 \begin{equation} 
{\rm E}(P)=\sigma\, \sqrt{\pi/2}
\end{equation}
\noindent and for the Rice distribution (presence of a source)
\begin{equation}
\begin{split}
{\rm E}(P)=\sigma\, \sqrt{\pi/2} \,\exp{(-P_0^2/4\sigma^2)}\,\big[(1+P_0^2/2\sigma^2)I_0(P_0^2/4\sigma^2)\\
+ (P_0^2/2\sigma^2) I_1(P_0^2/4\sigma^2)] 
\end{split}
\end{equation}
\noindent where $I_1$ is the modified Bessel function of first order. Note the difference between $E(P)$, the expectation value of the distribution, and $\left\langle P_0 \right\rangle $, the mean source polarisation of our sample. It is clear that the expectation of $P$ is not $P_0$, but a complicated expression involving $P_0$ and $\sigma$. This will introduce a significant bias when we try to determine $P_0$ by using the stacking technique.

There is also an analytic expression for the variance in both cases:
\begin{equation}
 {\rm var(P)}=(4-\pi)\sigma^2/2
\end{equation}
\noindent for the Rayleigh distribution and 
\begin{equation}
 {\rm var(P)}=2\sigma^2+P_0^2- {\rm E}(P)^2   
\end{equation}
\noindent for the Rice distribution, with ${\rm E}(P)$ given by Eq.$(3)$. See \cite{ARG09} for more details.

When we use the technique of stacking, we add up all the patches to obtain a final map. In this way, we enhance the signal-to noise ratio. The polarisation at the centre of this final stacked patch is
\begin{equation}
 P_s=\Sigma_{k=1}^n P_k 
\end{equation}
\noindent with $P_k$ the polarisation at the centre of the corresponding $kth$ patch. We also remove a residual background by subtracting the average of the fluxes outside the sources (those at least $3\sigma_{\rm beam}$ away from the central pixel) from the final map. 
Although by subtracting the off-source background we are slightly increasing the dispersion around the measured mean, as explained before, this is done to take into account the presence of residual foregrounds in our {\it Planck} maps in total intensity estimations and, therefore, in the patches centred at the positions of our sources (contrary to radio maps at low frequency, where these foregrounds are generally negligible). Calling $P_{0k}$ the polarisation amplitude of the source at the centre of the $kth$ patch, and taking into account the subtraction of the residual background, we find the following expressions for the expectation of the polarisation at the central pixel of the final stacked map and for its variance:
\begin{equation}
\begin{split}
 {\rm E}(P_s)=\Sigma_{k=1}^n \sigma\, \sqrt{\pi/2} \big(\,\exp{(-P_{0k}^2/4\sigma^2)}\,\big[(1+P_{0k}^2/2\sigma^2)I_0\\
 (P_{0k}^2/4\sigma^2)+ P_{0k}^2/2\sigma^2 I_1(P_{0k}^2/4\sigma^2)]-1\big) 
 \end{split}
\end{equation}
\begin{equation}
{\rm var}(P_s)=2n\sigma^2+\Sigma_{k=1}^n (P_{0k}^2-{\rm E}(P_k)^2) + n(4-\pi)\sigma^2/2p
\end{equation}
\noindent where p is the number of independent beams used for the calculation of the average background. Let us suppose, for the sake of simplicity, that all the sources have the same amplitude $ P_0=m\sigma$. In that case, we can easily compute the relative error $({\rm E}(P_s)/n-\left\langle P_0\right\rangle)/\left\langle P_0\right\rangle $ that we will make when estimating $\left\langle P_0 \right\rangle$ with $P_s/n$. For instance, for $m=1,2,3,4,5$ we obtain $ -0.7047 \,,  -0.4905 \,,   -0.3602  \,, -0.2815 \,,  -0.2304 $, respectively. As we can see, the bias is very significant.
  
If we carry out a  {\it quadratic stacking} defining $P_{qs}=\Sigma_{k=1}^n P_k^2$ and calculate ${\rm E}(P_{qs}/n)$, we find
 \begin{equation}
 {\rm E}(P_{qs}/n)=2\sigma^2+\left\langle P_0^2\right\rangle 
 \end{equation}
 \noindent where $\left\langle P_0^2\right\rangle $ is the average of $P_0^2$ for our sample of sources. Therefore, we can determine, in theory, $\left\langle P_0^2\right\rangle $ as
  \begin{equation}
 \left\langle P_0^2\right\rangle = {\rm E}(P_{qs}/n)-2\sigma^2 
  \end{equation}
 Besides, the r.m.s deviation  of  $P_{qs}/n$ can also be readily obtained
  \begin{equation}
 {\rm var}(P_{qs}/n)=(2\sigma/\sqrt{n})^2\,(\sigma^2+\left\langle P_0^2\right\rangle )
  \end{equation}
 Assuming, for simplicity, that all the sources have the same amplitude $P_0=m\sigma$ with $m=1,2,3,4,5$, we find the following values for the r.m.s. of $P_{qs}/n$ in terms of  $\sigma^2/\sqrt{n}: $ $      2.8284 \,,      4.4721\,,       6.3246\,,       8.2460 \,,        10.1978 $. Thus, we can obtain a good estimation of $\left\langle P_0^2\right\rangle $ with a low r.m.s. deviation (see Section \ref{sec:results}).
  
The knowledge of $\left\langle P_0^2\right\rangle  $ allows us to estimate $\left\langle \Pi^2\right\rangle$, an important quantity that can be found in the literature  for other samples \citep{TUC12} and that is used in Section \ref{sec:results} to characterize the log-normal distribution for $\Pi$.

This calculation cannot be directly applied to our case because the values of $P_{0k}$ will be different for each source and, since the sources are not detected in the polarisation maps, the individual information of each source about the S/N is not available. However, we can carry out simulations with injected sources and estimate the bias, as described in Section \ref{sec:s_injection}.

\subsection{Source injection}
\label{sec:s_injection} 

In order to investigate the robustness of stacking in our analysis, we perform a source injection test. We injected simulated compact sources in our real maps, at random positions but avoiding the position of real sources. In this way we can assess how well the method works in presence of potential systematics such as foreground contamination, noise bias and leakage.

In the sky region outside the Galactic mask, the sources in total intensity are simulated at each frequency independently following the model by \cite{TUC11}, with a flux limit of $\sim 316$ mJy (at 30 GHz). 

As demonstrated in \cite{STI14}, the assumed statistical distribution for the flux densities/degree of polarisation has a direct effect on the measured mean polarisation properties. In their work, they conclude that for their case, the best statistical distribution to be used is the one estimated directly from the detected sources of their sample. In our case, we are in a better situation because we have accurate observations of our same population of sources (for faintest sources and at slightly lower frequencies) by \cite{MAS13}: they observed with the Australia Telescope Compact Array during a dedicated, high sensitivity run ($\sigma_P \sim 1 mJy$) a 99 per cent complete sample of extragalactic sources with $S_{20GHz} > 500 mJy$, with declination $\delta < -30^{\circ}$ and a detection rate in polarisation at 20 GHz of 91.4 per cent. With this sample they were able to determine that the degree of polarisation of their sources follows a log-normal distribution. 
Therefore, considering that our compact radio sources are from the same parent population of those by \cite{MAS13}, we can safely apply their conclusions to our sample and assume a log-normal distribution without any further analysis.

In view of the above, the flux densities in polarisation are then simulated following a log-normal distribution for different values of $\langle \Pi \rangle$ in each simulation, and we assume a uniform random polarisation angle to compute $Q$ and $U$ for each source. From the simulated catalogue we then create the simulated sources map and convolve it with the FWHM of the instrument (which is different for each \textit{Planck} channel) before adding it to the real $Q$ and $U$ maps. Finally, we construct the new $P$ map (real $P$ map plus injected sources) to be used in the same stacking procedure as for the real data.

As explained in Section \ref{sec:noisebias}, we need to correct our measurements from the noise bias. In order to estimate this correction, we inject simulated sources in the real maps as described above. For each simulation, we vary randomly both the {\it location}, $\mu$, and {\it scale}, $\sigma$, parameters of the log-normal distribution (see Section \ref{sec:lognorm}). At each frequency channel, we limit the range of the simulated ($\mu$, $\sigma$) pairs by using the first tens of simulations and being guided by the measured $\left\langle \Pi  \right\rangle$ values, not yet bias corrected. As a general outcome, low $\sigma$ parameters produce Gaussian-like values around the $\mu$ parameter. On the contrary, high $\sigma$ values create a distribution with a long tail at the high end and, as a consequence, a mean value greater than the $\mu$ parameter. In other words, if two simulations have the same location parameter, the one with the bigger scale parameter will produce more sources with high polarisation degree and therefore higher recovered values.  

Then, we apply the same methodology described in Section \ref{sec:stacking} to obtain a recovered (biased) value. We repeat this procedure for at least 100 simulations at each frequency. By comparing the theoretical simulated $\left\langle \Pi \right\rangle = \left\langle P_s/S_s \right\rangle$ with the recovered values we are able to obtain the noise bias correction relationships shown as blue points in Figures \ref{fig:ex_res}, \ref{fig:gal_res}, \ref{fig:gal_res30} and \ref{fig:ex_res_pi2}. These blue points are simply the binned values, equally spaced in the x-axis, of the 100 individual simulations (shown as grey points, as an example, at 44 GHz for Figures \ref{fig:ex_res} and \ref{fig:gal_res}).  Please notice that most of the dispersion is caused by the brightest sources in each simulation that can modify greatly the input and/or recovered  values. Finally, we estimate our debiased measurements (red points) using a linear interpolation by means of a fit to the blue points of the derived noise bias correction relationships.

As for the simulations in the sky region inside the Galactic mask, it should be noticed that there is no source number counts model consistently describing the compact source populations inside this mask. However, we checked in the PCCS \citep{PLA13pccs} and in the PCCS2 that the model by \cite{TUC11} can still be considered a fair representation for the shape of the rough number counts for compact objects inside the adopted mask, at least for the purpose of estimating the noise bias correction with source injection. For this reason, we decide to apply the same procedure used for compact radio sources that lay outside the mask to estimate and correct the bias.

As described in Section \ref{sec:noisebias}, $\left\langle \Pi^2 \right\rangle $ should suffer of a lower contamination due to the noise bias that should be mainly taken into account by simply subtracting the mean in the stacking procedure. In any case, we compute the correction to this quantity in the same way as for $\left\langle \Pi \right\rangle $.

Moreover, {\it Planck} maps seem to suffer from a leakage between total intensity and polarisation \citep{PLAleak}. Whereas for the HFI channels this effect is negligible when estimating the polarized flux density of compact sources \citep[see][]{PLA15pccs2}, this might not be the case for the LFI ones. We use source injection to check its importance. In view of the above, we simulate the leakage of total intensity into polarisation as a normal distribution with a dispersion of 0.1 for each injected source and we apply the same stacking procedure. We found that even down to the $1\sigma$ level, the leakage is not affecting our results.

Finally, to check that our findings are model independent, we also perform simulations adopting the source number count model by \cite{DEZ05}. Comparing both sets of results we obtain that they are compatible within 1$\sigma$ and, thus, we can be reasonably confident in our conclusions.

\subsection{Log-normal distribution parameters }
\label{sec:lognorm} 

As discussed in Section \ref{sec:s_injection}, we can safely adopt a log-normal distribution when simulating the flux densities in polarisation. If we assume a log-normal distribution for $\Pi$, the PDF of which is given by
 \begin{equation}
 f\left( \Pi;\mu,\sigma  \right) =\dfrac{1}{\Pi \sigma \sqrt{2\pi}} \exp\left( -\dfrac{(\ln(\Pi) - \mu)^2}{2\sigma^2}\right) 
 \end{equation}
we use $\left\langle \Pi \right\rangle $ and $\left\langle \Pi^2 \right\rangle $ for the determination of $\mu$ and $\sigma$. Following \cite{LOGNORM}, they are given by 
 \begin{equation}
 \mu=\ln\left( \dfrac{\left\langle \Pi \right\rangle^2}{\sqrt{\left\langle \Pi^2 \right\rangle}}  \right) 
 \end{equation}
  \begin{equation}
\sigma = \sqrt{\ln \left(  \dfrac{\left\langle \Pi^2 \right\rangle}{\left\langle \Pi \right\rangle^2} \right) }
 \end{equation}
 and their errors are, respectively
 \begin{equation}
{\rm var}(\mu) = \dfrac{4}{\left\langle \Pi \right\rangle^2} {\rm var}(\left\langle \Pi \right\rangle) + \dfrac{1}{\left\langle \Pi^2 \right\rangle} {\rm var}(\sqrt{\left\langle \Pi^2 \right\rangle)}
 \end{equation}
 \begin{equation}
{\rm var}(\sigma) = \dfrac{1}{\sigma^2} \left(  \dfrac{{\rm var}(\left\langle \Pi \right\rangle)}{\left\langle \Pi \right\rangle^2} + \dfrac{1}{\left\langle \Pi^2 \right\rangle} {\rm var}(\sqrt{\left\langle \Pi^2 \right\rangle)}  \right) 
 \end{equation}
 
 From $\mu$ we can then compute the median fractional polarisation as 
  \begin{equation}
\Pi_m=\exp{(\mu)}
 \end{equation}

Therefore, from the recovered values of $\langle \Pi \rangle$ and $\langle \Pi^2 \rangle$ we can compute the parameters $\mu$ and $\sigma$ characterizing the adopted distribution.

\section{Results}
\label{sec:results} 

\begin{figure*}
 \centering
 \includegraphics[width=16.0cm]{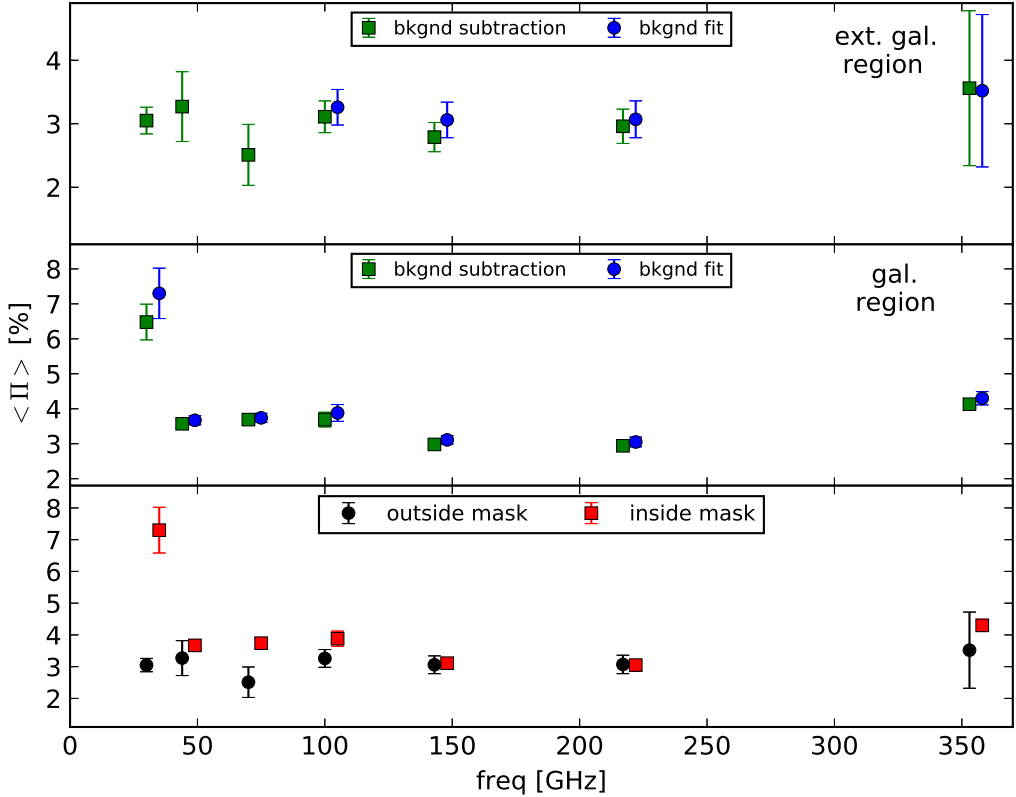}
 \caption{Mean fractional polarisation for each frequency outside (top) and inside (center) the \textit{Planck} Galactic mask with $f_{sky}$=60 per cent. The green squares are obtained with the flat background estimation, the blue circles with the background fit as described in the text. The bottom panel compares the results obtained outside (black circles) and inside (red squares) the mask.}
 \label{fig:polfrac}
\end{figure*}

\begin{figure*}
 \centering
 \includegraphics[width=16.0cm]{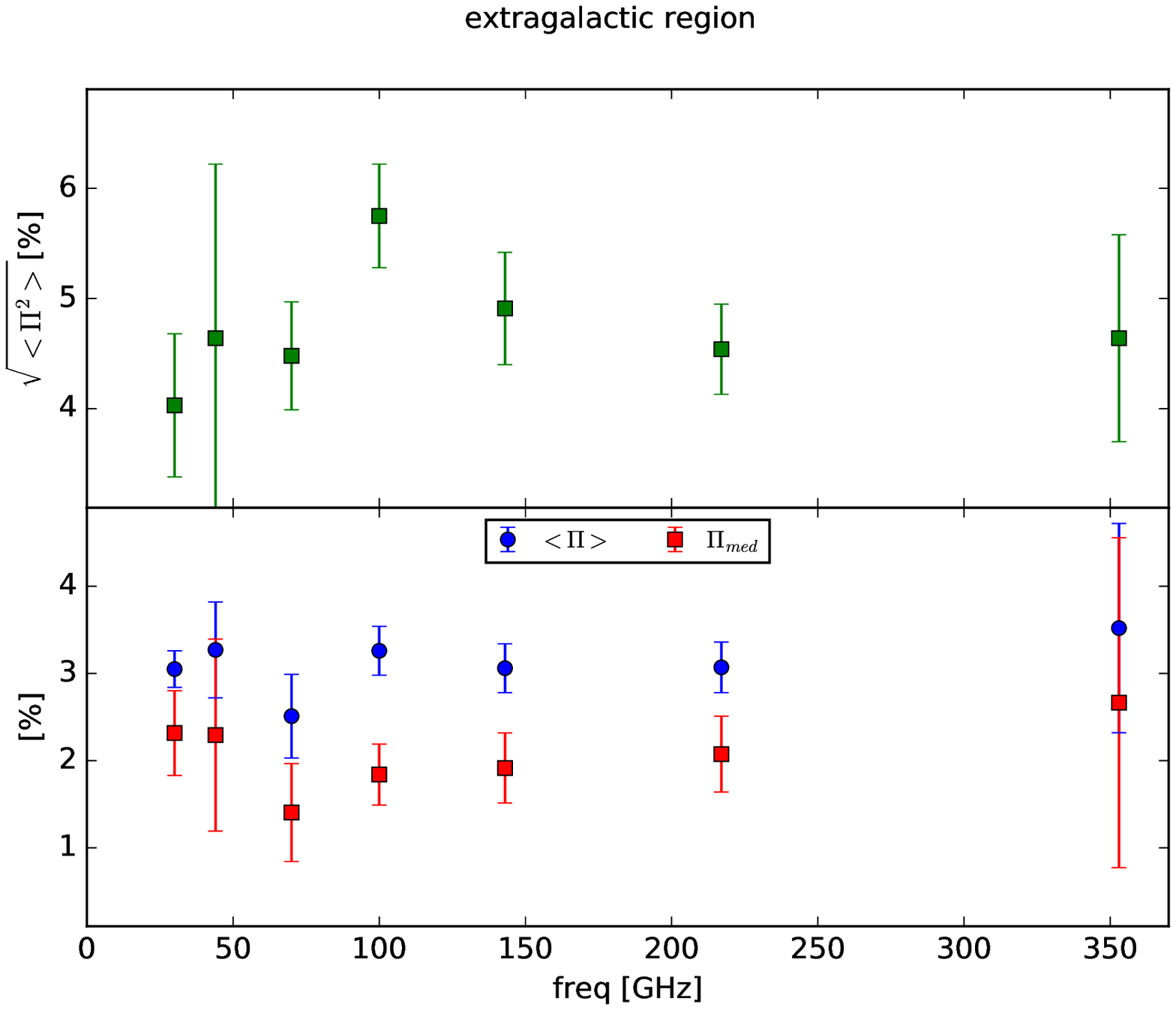}
 \caption{Top panel: $\sqrt{\langle \Pi^2 \rangle}$ for the extragalactic region of the sky (green squares). Bottom panel: comparison between the mean fractional polarisation (blue circles) and the median fractional polarisation (red squares).}
 \label{fig:polfrac_pi2}
\end{figure*}

Of the 1560 sources in our sample, 881 are outside the {\it Planck} GAL60 (plus the MCs regions) mask and 679 are inside it. In both cases we have a fairly good sample. In these positions we perform stacking in the \textit{Planck} channels from 30 to 353 GHz both in total intensity and polarisation. From the patches resulting from the stacking, we estimate $\langle S \rangle$,  $\langle P_0 \rangle$, $\langle \Pi \rangle$ and $\langle S^2 \rangle$,  $\langle P_0^2 \rangle$, $\langle \Pi^2 \rangle$ and we compute the errors from the standard deviation of the residual background fluctuations,  as described in  Section \ref{sec:stacking}. The results are summarized in the left part of Tables~\ref{tab:polfrac_ext} and \ref{tab:polfrac_gal}.

To estimate $\left\langle \Pi  \right\rangle $ from the bias-uncorrected values, we use the procedure described in Section \ref{sec:s_injection}, whose results are shown in Fig. \ref{fig:ex_res} for the extragalactic region and in Fig. \ref{fig:gal_res} for the Galactic region of the sky. The grey points in the 44 GHz panel are the 100 simulations (the inputs are on the y-axis and the recovered values with stacking are on the x-axis). For all the panels, the blue points are obtained by binning the simulations results with a binning step of about 0.5 in the y-axis. The red squares are the values recovered with data, namely performing stacking in the PCCS2 30 GHz positions: these results are computed by performing a linear interpolation to the values obtained with simulations and using it to correct/debias the measured $\langle \Pi \rangle$ (see Section \ref{sec:noisebias}). 

Due to the stronger foreground contamination inside the Galactic region at 30 GHz (synchrotron emission), our individual measurements of the simulated data suffer from a much larger dispersion. The results at this frequency are shown in Fig.  \ref{fig:gal_res30}. As described in more detail in Section \ref{sec:comparison}, at 30 GHz we are able to perform a direct check of our results, i.e. our correction procedure, that confirms that even with such high level of measurement dispersion we are still able to get acceptable results.

In the extragalactic region of the sky, the noise bias correction for $\langle \Pi \rangle$ goes approximately from 3 to 6, which corresponds to compact sources with a polarisation value that presents a S/N of about 0.5-1.5, as discussed in Section \ref{sec:noisebias}. These S/N levels are the expected ones, since most of the sources are not directly detected in polarisation with {\it Planck}. The same approach as decribed in Section \ref{sec:s_injection} has been applied to correct also the values obtained for $\langle \Pi^2 \rangle$ (in the extragalactic region of the sky; see Fig. \ref{fig:ex_res_pi2}). 

As anticipated in \ref{sec:noisebias} the noise bias correction for $\sqrt{\langle \Pi^2 \rangle}$ is much lower than for the $\langle \Pi \rangle$ case: we find a mean value of 1.2 to be compared with 4.1 for $\langle \Pi \rangle$, in the extragalactic region of the sky.

Figs. \ref{fig:polfrac} and \ref{fig:polfrac_pi2} are a summary of the noise bias corrected results on $\langle \Pi \rangle$ and $\sqrt{\langle \Pi^2 \rangle}$ we obtain with stacking.

Fig. \ref{fig:polfrac} shows the estimated $\langle \Pi \rangle$ for the different cases. In the top panel (sky region outside the Galactic mask) we compare the results obtained by subtracting the estimation of the mean value of the background from the source flux density in total intensity (green squares) with those obtained by subtracting the model of the background, as described in Section \ref{sec:stacking} (blue circles). In both cases the results are compatible within one sigma level. From the results in the extragalactic region, it can be concluded that all the estimated values of  $\langle \Pi \rangle$ are almost consistent within one sigma for all the channels (with a weighted mean value of 3.08 per cent).

The central panel shows the same comparison, but for the sky region inside the adopted mask. Again, the results obtained by modelling the background (blue circles) or directly estimating its mean value (green squares) are consistent with each other within the errors. $\langle \Pi \rangle$ seems to remain constant across the range of frequencies from 44 to 353 GHz (with a weighted mean equal to 3.51 per cent), while at 30 GHz it reaches much higher values (7.30 $\pm$ 0.72) per cent.

Finally, in the bottom panel we compare our results inside (red squares) and outside (black circles) the Galactic mask. With the exception of the 30 GHz channel, there is no relevant statistical difference between the values obtained inside and outside the adopted mask.

In the top panel of Fig. \ref{fig:polfrac_pi2} we show the results for $\sqrt{\langle \Pi^2 \rangle}$ at each frequency for the extragalactic region of the sky. In both figures we show the $1\sigma$ error resulting from the interpolation procedure.

As described in Section \ref{sec:lognorm}, from the recovered values of $\langle \Pi \rangle$ and $\langle \Pi^2 \rangle$ we compute the parameters $\mu$ and $\sigma$ for the log-normal distribution in the extragalactic region and they are summarized in Table \ref{tab:polfrac_ext}. 

In the bottom panel of Fig. \ref{fig:polfrac_pi2}  we compare the median and the mean values for the estimated polarisation fraction. As expected for a log-normal distribution, the values for the median are lower than those for the mean in the region outside the Galactic mask.

The foreground fluctuations inside the the adopted Galactic mask affect more severely the $\sqrt{\langle \Pi^2 \rangle}$ measurements producing huge dispersion between the different simulations. For this reason we decided not to attempt any $\sqrt{ \langle\Pi^2 \rangle}$ measurements or noise bias correction and, as a consequence, any estimation of the $\mu$ and $\sigma$ parameters of the adopted log-normal distribution.

\subsection{Comparison with results based on detected extragalactic compact sources}
\label{sec:comparison}

At lower frequencies, \citet{MAS08} -- by exploiting the {\it Bright Source Sample} which consists of sources with $S_{\rm 20GHz} > 500$ mJy selected in the AT20G survey \citep{MUR10} and observed with the Australia Telescope Compact Array -- found a $\langle \Pi \rangle$ value of 2.5 per cent at $\sim$20 GHz and a slight increase of $\Pi$ with frequency (between 4.8 and 20 GHz) and little correlation between polarized and total intensity spectra. Moreover, \cite{MAS13} with sources selected again from the AT20G survey and with $S_{\rm 20GHz} > 500$ mJy found in the same range of frequencies no significant evidence of correlation between $\Pi$ and frequency or $\Pi$ and total flux density and they recovered a value for $\langle \Pi \rangle$ of 2.79 per cent at 18 GHz. This value is close to the one we found in the extragalactic region at 30 GHz (3.05 $\pm$ 0.21) per cent.
\cite{SAJ11} observed with the Very Large Array, for frequencies from 4.86 to 43.34 GHz, typical values for $\langle \Pi \rangle$ of  2 - 5 per cent  in a sample of sources selected in the AT20G catalogue with $S_{\rm 20GHz}>40$ mJy.
More recently, \cite{GAL17} found a median value for the polarisation fraction of 2.01 per cent at 33 GHz for their sample of 53 sources selected from the {\it faint PACO sample} \citep{BON11}, with $S_{\rm 20GHz} > 200$ mJy. They also found no trend with frequency or flux densities (up to 38 GHz) for $\Pi$, in agreement with our current findings.

At 30 GHz we perform a test with the PCCS2 catalogue to check that our noise bias correction is properly taken into account as described in Section \ref{sec:noisebias}. We compute $\langle \Pi \rangle$ for a subsample of our sources that are detected in polarisation and have $P > 2P_{\rm err}$ where $P$ is the polarized flux density and $P_{\rm err}$ is the error measurement in the PCCS2 (this subsample consists of 53 sources). We recover a value for the mean fractional polarisation $\langle \Pi \rangle$ of (7.56 $\pm$ 0.53) per cent (green diamond in Fig. \ref{fig:gal_res30}), in agreement with the value computed with the PCCS2 official catalogue for the same subample: (7.20 $\pm$ 0.15) per cent (green band). It is worth mentioning that this subsample is mostly made up of sources inside the Galactic mask and its $\langle \Pi \rangle$ value is very close to the one found for our sample inside the adopted mask (7.30 $\pm$ 0.72) per cent implying that probably the brightest sources in polarisation in the Galactic plane are also the ones with the higher mean fractional polarisation.

For frequencies higher than 40 GHz, \cite{TUC12} gave a preliminary estimate of $\sqrt{\langle \Pi^2 \rangle}$: 3.5-4.6 per cent for FRSQ and 4.2-5.5 per cent for BL Lac, in general agreement with our findings. This agreement is not surprising, since this quantity depends on the tail of the distribution for $\Pi$, where most of the sources included in the catalogues used by these authors usually lie. On the other hand, the values of $\Pi_m$ estimated by \cite{TUC12} are a bit higher than ours (2.5-3.6 per cent for FSRQ and 3.0-4.3 per cent for BL Lac), probably due to the fact that our measurements are based on fainter sources (in polarisation) with respect to the ones in the catalogues used by them.

Moreover, \cite{LOP09} found lower $\langle \Pi \rangle$ values for WMAP 5-yr data (1.7, 0.91, 0.68, 1.3 per cent from 23 to 61 GHz, respectively).  For the PCCS2 we compute the following values for $\langle \Pi \rangle$: 5.6, 7.9, 7.8, 6.5, 8.2 and 12.2 per cent at 30, 44, 70, 100, 143 and 217 respectively, in agreement with Fig. 5 in \cite{PLA15pccs2}. 
These higher values of $\langle \Pi \rangle$, directly calculated from the PCCS2\footnote{In this case we have used {\it all} the sources present in the PCCS2 for which a flux in total polarization, $P$, was measured. On the contrary, the estimate displayed by a green diamond in Fig. \ref{fig:gal_res30} is calculated only from PCCS2 sources for which $P>2P_{err}$).}, can be explained by taking into account that these same values are not estimated from a complete sample but from sources that can be detected at only one particular frequency and are, thus, biased towards sources with higher polarisation.

On the contrary, the stacking analysis allows us to reduce the background noise and enhance the signal. In this way we are able to compute $\langle \Pi \rangle$ at different frequencies for the {\it same} sample of sources, identified by their positions, regardless of whether they can be detected or not. Moreover, we also make a distinction between Galactic and extragalactic regions of the sky.

Concerning the values we found for the log-normal distribution of $\Pi$, we can compare our 30 GHz results with those by \cite{MAS13} at 18 GHz. They found a value for $\sigma$ of 0.90 and a median polarisation fraction $\Pi_m$ of 2.14 per cent. They are in good agreement with our findings: $\sigma$=(0.7$\pm$0.2) and $\Pi_m$=(2.2$\pm$0.4) per cent (given by $\Pi_m=\exp{(\mu)}$).

\section{Conclusions}
In this work we apply stacking to estimate the mean fractional polarisation of radio sources at {\it Planck} frequencies. Our sample consist of 1560 sources, i.e. the sources in the PCCS2 catalogue at 30 GHz. They are  divided into two subsamples: 679 inside and 881 outside the $Planck$ Galactic mask with $f_{\rm sky} = 60$ per cent. 
At higher {\it Planck} frequencies the positions in the sky corresponding to these sources are used.
$\langle \Pi \rangle$ is computed at 30, 44, 70, 100, 143, 217 and 353 GHz in the {\it Planck} maps. 

Due to detection bias, we perform background subtraction to compute the mean flux density in total intensity for the sources in two different ways (see Section \ref{sec:stacking}): by subtracting the mean value obtained in the region of the patch $3\sigma_{\rm beam}$ away from its the center and by modelling the background before its subtraction. In both cases we find compatible results (within their errors).

When we use the stacking technique to enhance the signal, we find that there is a significant bias in the determination of the average source polarisation (called noise bias, see Section \ref{sec:noisebias}). In order to estimate and correct for this bias, we perform simulations with injected sources (Section \ref{sec:s_injection}).

We also perform simulations by following two different source counts models \citep{TUC11,DEZ05}. The results are compatible at the $1\sigma$ level.

For the sky region outside the Galactic mask, we obtain values for $\langle \Pi \rangle$ in the different channels that are not too different taking into account their errors: they go from the lowest value of (2.51$\pm$ 0.48) per cent at 70 GHz up to (3.52 $\pm$ 1.20) per cent at 217 GHz. Remarkably, there is a good agreement with low-frequency survey results: 2.5 per cent by \cite{MAS08}, 2.79 per cent by \cite{MAS13} and 2-5 per cent by \cite{SAJ11}.

We also estimate the parameters characterizing the log-normal distribution for $\Pi$: we obtain a weighted mean for $\mu$ over the whole {\it Planck} frequency range of 0.7 (that would imply a median value for $\Pi$ of 1.9 per cent, lower than the measured mean value of 3.08 per cent, as expected for a lognormal distribution) and a weighted mean for $\sigma$ of 1.0.

For the region inside the Galactic mask, the behaviour is similar to the previous case with a tendency to slightly higher values. $\langle \Pi \rangle$ ranges from a minimum value of (3.05 $\pm$ 0.14) per cent at 217 GHz to a maximum of (4.30 $\pm$ 0.19) per cent at 353 GHz,  with the exception of the 30 GHz channel, where $\langle \Pi \rangle$ is much higher, reaching a value of (7.30 $\pm$ 0.72) per cent.

This excess of polarisation is not seen in our source injection test and therefore indicates a stronger emission in polarisation that, apparently, only affects frequencies lower than 30 GHz in the Galactic region. The fact that the measured mean of the full sub-sample inside the Galactic region is very similar to the value measured considering only the sources detected  at 30 GHz in polarisation, implies that these few very bright sources completely dominate the signal. Therefore, the value measured directly from the PCCS2 cannot be considered a correct statistical representation of the  sub-sample properties.

The strong contamination produced by the diffuse emission of our Galaxy inside the Galactic region does not allow us to estimate $\sqrt{\langle \Pi^2 \rangle}$ and its bias corrections and, therefore, the log-normal distribution parameters. Considering these limitations and taking into account the approximations adopted to simulate the source population in this sky region, it is clear that the Galactic region results are less robust than the extragalactic region ones, but of great interest in any case.

In conclusion, the stacking method can be useful to estimate the mean and median values of the fractional polarisation of mainly undetected compact sources by exploiting the knowledge of their positions in the sky. 
At the lowest {\it Planck} frequencies and at high Galactic latitudes our present results are in general agreement with other published works on extragalactic radio sources \citep[][]{SAJ11,MAS13,GAL17}. At higher frequencies we are presenting novel results that could be useful in estimating the polarized source number counts and, consequently, the contamination due to these sources for the detection of the primordial CMB E and B mode polarisation. Therefore, our results are very useful when planning future CMB experiments in polarisation such as $CORE$ \citep{CORE}.

\section*{Acknowledgements}
We thank the anonymous referee for the insightful comments and remarks that helped a lot in improving the original paper and for the careful reading of the manuscript. The authors also thank M. L\'{o}pez-Caniego and B. Partridge for useful comments.

The authors acknowledge financial support from the I+D 2015 project AYA2015-65887-P (MINECO/FEDER). J.G.N also acknowledges financial support from the Spanish MINECO for a 'Ramon y Cajal' fellowship (RYC-2013-13256).



\bsp	
\label{lastpage}
\end{document}